\documentstyle[editedvolume]{crckapb} 

 \newcommand{\ts}{\thinspace} \newcommand{\etal}{et
al.}
\def\boxit#1{\vbox{\hrule\hbox{\vrule\kern3pt\vbox{\kern3pt#1\kern3pt}
\kern3pt\vrule}\hrule}}
\def\ga{\mathrel{\hbox{\raise0.3ex\hbox{$>$}\kern-0.8em\lower0.8ex
\hbox{$\sim$}}}}
\def\la{\mathrel{\hbox{\raise0.3ex\hbox{$<$}\kern-0.75em\lower0.8ex
\hbox{$\sim$}}}}

\begin{opening}
\title{THE ``GREAT DEBATE'':\protect\\
       THE CASE FOR AGNs}

\subtitle{On behalf of the ``green team''}

\author{D.B. SANDERS}
\institute{Institute for Astronomy, University of Hawaii\\
           2680 Woodlawn Drive, Honolulu, HI 96822 USA}

\end{opening}

\runningtitle{THE CASE FOR AGNs}

\begin{document}

\begin{abstract}
We summarize the evidence from multiwavelength observations that the dominant
power source in the majority of ultraluminous infrared galaxies (ULIGs) may be
an active galactic nucleus (AGN).  In the broader context of the debate, we
also show that --- 1.\ ULIGs are indeed a key stage in the transformation of
merging gas-rich disks into ellipticals, \ 2.\ ULIGs are plausibly the
precursors of quasi-stellar objects (QSOs), \ and\ 3.\ ULIGs do appear to be
local templates of the high luminosity tail of major gas-rich mergers at
$z{\ts}\sim${\ts}1--4.  \end{abstract}

\section{Background}

The nature of the dominant power source in ultraluminous infrared galaxies
(hereafter ULIGs) has been the subject of intense debate, ever since their
discovery in significant numbers by the all-sky survey carried out by the
Infrared Astronomical Satellite ({\it IRAS}).  The debate has intensified once
again following recent mid- and far-infrared spectroscopic observations of
ULIGs by the Infrared Space Observatory ({\it ISO}), and is the main theme of
this workshop.  The scientific organizing committee (SOC) asked the conference
participants to form two teams (with meeting rooms in the ``blue tower" and the
``green tower") charged with marshaling the evidence in favor of starbursts and
AGNs respectively, in preparation for ending the conference with a ``Great
Debate".  This article presents the evidence assembled by the ``green team" as
to the fraction of the total bolometric luminosity of ULIGs that can reasonably
be attributed to dust-enshrouded AGNs, and answers additional questions posed
by the SOC on the relevance of ULIGs to galaxy transformations.

\subsection{KEY TOPICS FOR THE DEBATE}

The SOC proposed the following four topics for the debate: 

\setbox4=\vbox{\hsize 28pc \noindent\hangindent=0.25in \strut 
{\it (1)\ Most of $> 10^{12} L_\odot$ ULIGs are predominantly
powered by ?\\  (\underbar{heavily dust enshrouded
AGN})\ or\ (\underbar{circumnuclear starbursts})}

\noindent\hangindent=0.25in {\it (2)\ ULIGs follow a merger sequence 
from colliding disk galaxies with large bulges to ellipticals.}

\noindent\hangindent=0.25in {\it (3)\ ULIGs are precursors of QSOs.}

\noindent\hangindent=0.25in {\it (4)\ ULIGs are local templates of 
the high luminosity tail of mergers at $z${\ts}={\ts}1--4.} \strut}

$$\boxit{\box4}$$

Much of the Green Tower discussion was devoted to Topic 1, which 
most participants had thought to be the main theme of the Workshop.  
Much of our time was spent focusing on the nearest and best studied 
ULIGs, and on comparing the observational evidence from across 
the electromagnetic spectrum for and against the presence of a 
dominant AGN.  Topic 3 seemed to blend naturally with the discussion 
of Topic 1.  

Discussion of Topic 2 drew heavily on several large optical and near-infrared
imaging studies of complete samples of low-$z$ {\it IRAS} galaxies, while Topic
4 relied on more recent studies of a limited number of high-$z$ {\it IRAS}
sources and sources recently identified in deep submillimeter images obtained
with the Submillimeter Common-User Bolometer Array (SCUBA: Holland et al. 1999)
on the James Clerk Maxwell Telescope (JCMT) on Mauna Kea.

\section{Opening Remarks}

The origin and evolution of ULIGs continues to be a subject of intense
research, and one that has taken on renewed importance following new
infrared/submillimeter results from {\it ISO} and SCUBA/JCMT, and from new
X-ray observations with {\it ROSAT}, {\it ASCA}, and {\it BeppoSax}.  Strong
evolution in the space density of luminous infrared galaxies has been detected
in the first deep mid-infrared surveys with ISOCAM (Taniguchi et al. 1997;
Aussel et al. 1999) and deep far-infrared surveys with ISOPHOT (Kawara et al.
1998; Puget et al. 1999).  Reports from the first deep submillimeter surveys
with SCUBA on the JCMT (Smail et al. 1997; Hughes et al. 1998; Barger et al.
1998; Eales et al. 1999) show that the space density of ULIGs at high redshift
($z > 1$) appears be sufficient to account for nearly all of the
far-infrared/submillimeter background radiation (e.g. Barger, Cowie, \& Sanders
1999), and depending on their exact redshift distribution, produces an infrared
luminosity density that exceeds that in the optical/UV by factors of 2--5 (e.g.
models by Blain et al. 1999). In addition, the discovery that much of the X-ray
background appears to be produced by a population of heavily obscured AGN (e.g.
Fabian \& Barcons 1992; Boyle et al. 1995; Almaini et al. 1998), objects which
have been largely missed in optical surveys due to extremely heavy obscuration
along the line of sight (e.g. $N_{\rm H} \ga 10^{24-25}${\ts}cm$^{-2}$:
Maiolino et al. 1998) has clearly renewed interest in studies of
infrared-selected AGN.

It seems clear that detailed observations of nearby ULIGs are a required first
step in order to better understand these objects, and in particular to unravel
the nature of the dominant energy source responsible for their enormous
infrared luminosity.  Given the evidence from millimeterwave interferometer
observations which indicate large absorbing columns toward the nuclei of all
ULIGs [typically $N({\rm H}_2) > 10^{24}${\ts}cm$^{-2}$: e.g. Bryant \&
Scoville 1996), it has always seemed prudent to use a broad multiwavelength
approach to study these sources, thus we welcome the new mid- and far-infrared
{\it ISO} data as the latest tool in the study of ULIGs.

The four topics posed for the debate are answered in the order they were 
posed.

\subsection{WORKING DEFINITIONS}

The following definitions have been in wide use, and are adopted for the 
purpose of this debate:

\vspace{0.1cm}

{\small 
$L_{\rm ir}$: \hskip 0.35in $L(8-1000{\ts}\mu m)$    

ULIG: \hskip 0.18in  $L_{\rm ir} \ge 10^{12}{\ts}L_\odot$\footnote{$L_{\rm ir}
\ge 10^{12} L_\odot$ is equivalent
	      to the minimum bolometric luminosity of a QSO, i.e. $M_{\rm B} <
	      -22.3$\ ---\
			  which is equivalent to $M_{\rm B} < -23$ (Schmidt \&
			  Green 1983) corrected for $H_{\rm o} =
			  75${\ts}km{\ts}s$^{-1}${\ts}Mpc$^{-1}$, $q_{\rm o} =
			  0$\ ---\  and a bolometric correction for QSOs of
			  $\sim{\ts}11.8 \times \nu L_\nu (0.43{\ts}\mu {\rm
			  m})$ (e.g. Elvis et al. 1994; see also Sanders \&
			  Mirabel 1996).}

\hangindent=0.83in ``warm'':  \hskip 0.05in $f_{25}/f_{60} > 0.2$ (as
originally defined by de Grijp et al. 1985; also Low et al. 1998, Sanders et
al. 1988b)

\hangindent=0.83in AGN: \hskip 0.23in  A compact nuclear region producing
energy by non-stellar processes. Also, a strong ``broad-line region" (BLR) with
doppler motions $> 2000${\ts}km{\ts}s$^{-1}$ (HWHM) from gas in a region
$<${\ts}1{\ts}pc in diameter.

RQQSO: \hskip 0.03in radio-quiet quasi-stellar object ($L_{20cm}/L_{1\mu m} \la
10^{-5}$) }

\section{``Ultraluminous Galaxies: Monsters or Babies''}

This was the advertised title of the Workshop, and with appropriate rewording,   
became the first topic for the debate:

\setbox4=\vbox{\hsize 28pc \noindent\hangindent=0.25in \strut 
{\bf Q1:\ Most of the $>${\ts}10$^{12}${\ts}$L_\odot$ ULIGs are predominantly 
 powered by AGN{\ts}?  or Starburst{\ts} ?} \strut}

$$\boxit{\box4}$$
 
\vspace{0.2cm}

We define the words ``most'' and ``predominantly'' in the way that they have
been used throughout this workshop, i.e. to mean  ``$>${\ts}50{\ts}\%''.

From the beginning it seemed clear that the AGN camp could relatively easily
reach a consensus on dominant AGNs for those ULIGs with ``warm'' mid-infrared
colors, nearly all of which have Seyfert-like optical and/or near-infrared
spectra, and for the most luminous ULIGs, i.e. the hyperluminous objects with
$L_{\rm ir} > 10^{13}${\ts}$L_\odot$, where {\it all} of the currently
identified objects indeed have ``warm'' colors and typically Seyfert{\ts}2
emission lines in direct optical emission (e.g. IRAS F09105+4108: Kleinmann \&
Keel 1987; IRAS F15307+3252: Cutri et al. 1994; IRAS F10214+4724:
Rowan-Robinson et al. 1991), but have been shown to contain hidden broad line
regions in polarized optical light (e.g. Hines et al. 1995), or in direct
near-infrared emission (e.g. Veilleux, Sanders, \& Kim 1997, 1999).

There was little or no agreement on the dominant energy source for the cooler
ULIGs which make up the bulk of the ULIG population by number.  As more
complete multiwavelength data sets were assembled for individual objects, what
was intriguing was the fact that for a significant fraction of the cool ULIGs,
different wavelength data often gave contradictory results; for example
relatively strong X-ray emission, or the clear presence of an AGN-like radio
core, while mid-infrared and optical line diagnostics favored starbursts.  What
was also apparent was that previous statements about AGN-like properties of
ULIGs often used as a benchmark objects which were radio-loud (e.g. 3C sources
such as 3C{\ts}273) when referring to the mean radio and/or X-ray properties of
AGN, despite the fact that the great majority of QSOs are {\it radio-quiet} and
relatively X-ray week.

It was decided that the best way to attack Topic 1 was an in depth study of the
five nearest ULIGs, objects for which a substantial body of high resolution,
multiwavelength data already exists, and then to compare the properties of
these objects with the mean properties of radio-quiet QSOs (RQQSOs).  Figure 1
presents images of the 5 nearest ULIGs and Table 1 summarizes the large scale
properties of each.

\begin{figure}[hbp] \includegraphics{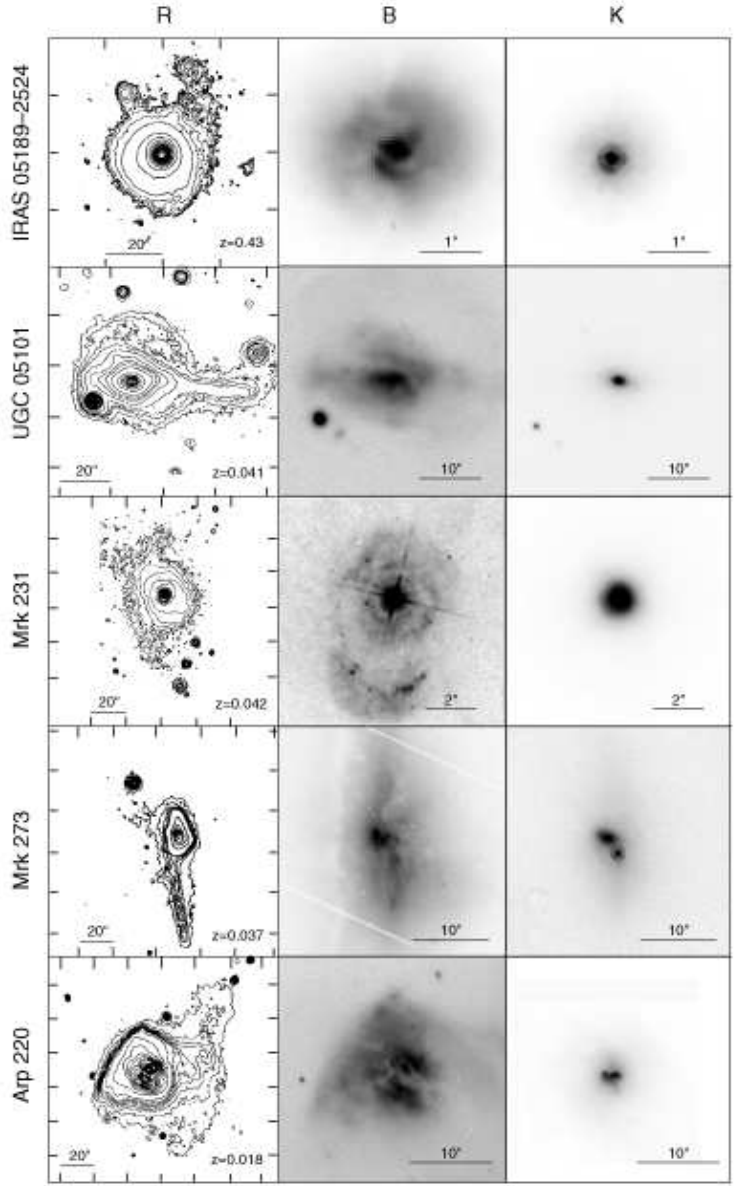} \vspace{16cm} \caption{The five nearest ULIGs
from the original {\it IRAS} BGS (Soifer et al. 1987).  The large scale R-band
images were obtained with the Palomar 5{\ts}m telescope (Sanders et al.
1988a,b,c). Tick marks represent intervals of 20$^{\prime\prime}$.  The
logarithmic stretch is designed to reveal both faint large scale features and
bright nuclei. The optical (B-band) {\it HST} images are from Surace et al.
(1998).  The K-band images are from  Surace et al. (1998) and Scoville et al.
(1999).  The plate scale for the B and K images is indicated in the lower right
corner of each panel.  The greyscale for the B and K images has been adjusted
to show only the brightest features.  For each object (row) all three
wavelength panels (R,B,K) are centered at the same (R.A., Dec).  } \end{figure}

\begin{figure}
\includegraphics{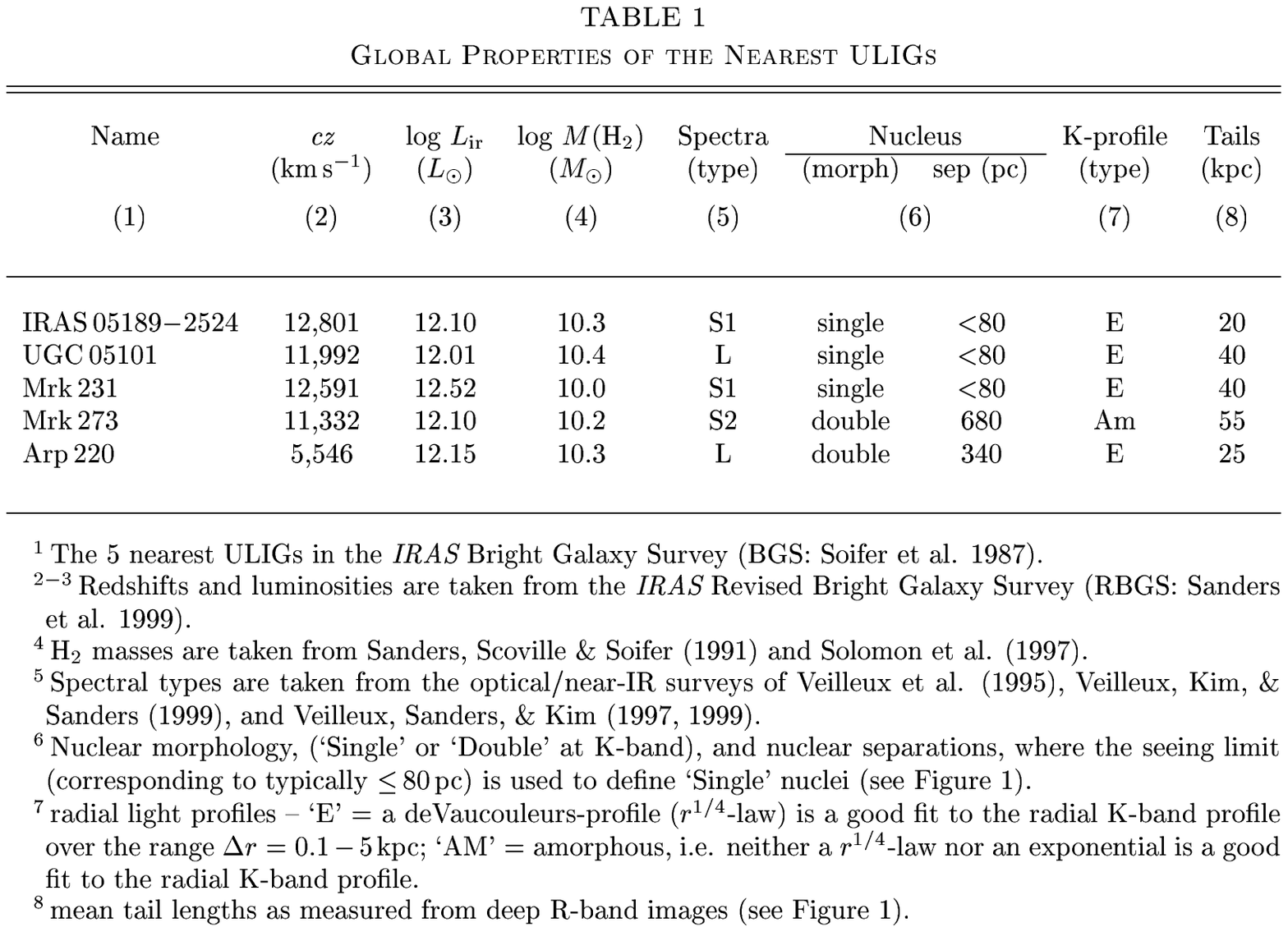}
\vspace{9.5cm}
\end{figure}

\subsection{Multiwavelength Properties of the Nearest ULIGs}  

Table 2 lists the radio--to--X-ray properties of the five nearest ULIGs,  
including the latest high-resolution radio data from the VLA and VLBA  
and X-ray data from ASCA.  In addition to the extensive notes included 
with Table 2, a brief summary of properties by wavelength band is given below. 

\begin{figure}
\includegraphics{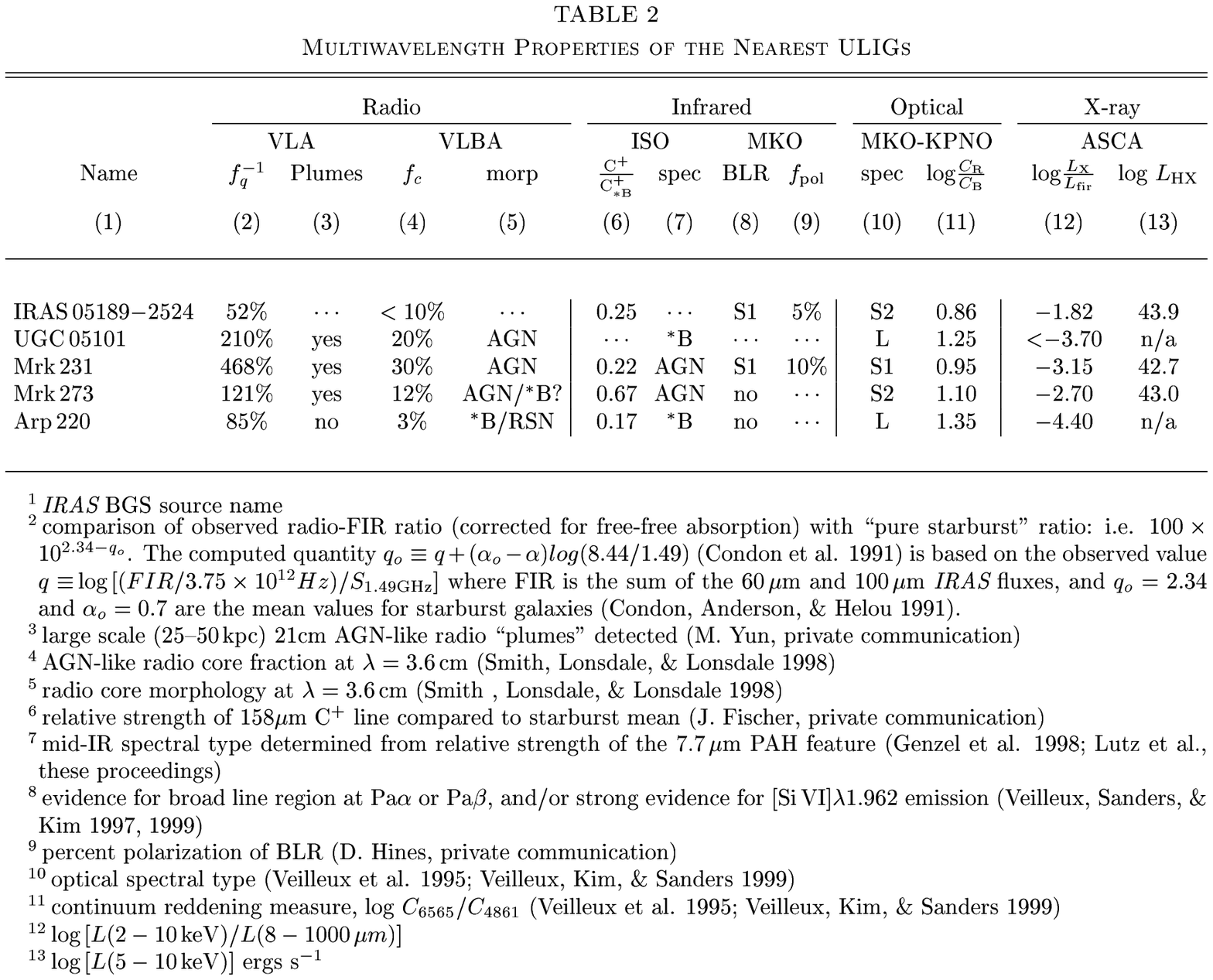}
\vspace{10cm}
\end{figure}

\subsubsection{Radio}

One measure commonly used to distinguish starbursts from AGN is the
FIR-to-radio correlation parameter $q_o$; starburst galaxies show a tight
correlation around a value of $q_o = 2.34$ ($\sigma \sim 0.1$), while
radio-loud AGN often have values as small as $q_o =${\ts}0 to -1.  The majority
of ULIGs have values of $q_o$ within the range $q_o = 2.0-2.6$, however {\it so
do the majority of RQQSOs}, which presents a problem when using a global
parameter such as $q_o$ to distinguish starbursts from {\it radio-quiet} AGN.

A more direct method of distinguishing AGN from starbursts at radio wavelengths
is to look for the presence or absence of a true high-brightness temperature
AGN-like radio core.  VLBA observations of RQQSOs find typical radio-AGN core
fractions in the range of 20--50{\ts}\% at 1.4{\ts}GHz (P. Barthel and H.
Smith, private communication).  Two of the five ULIGs in Table 2 (Mrk{\ts}231
and UGC{\ts}05101) have AGN-core strengths within this range; perhaps not
surprisingly, these two objects also have the smallest $q_o$ values.

A somewhat unexpected finding is the detection with the VLA of large scale 
(i.e. 20--50{\ts}kpc) radio ``plumes'' in three of the five ULIGs 
(UGC{\ts}05101, Mrk{\ts}231, Mrk{\ts}273), similar to the bipolar plumes 
or extended jets seen in radio galaxies (M. Yun, this workshop) suggesting 
the presence of a powerful radio-AGN core in these objects. 

\subsubsection{Mid-IR}

The mid-infrared (i.e. $\lambda${\ts}$\sim${\ts}5--50{\ts}$\mu$m) begins to
directly probe the dominant peak of the spectral energy distributions (SEDs) of
ULIGs.  The most promising new diagnostic tool in this wavelength range is that
provided by {\it ISO} spectroscopy.  Genzel et al. (1998), Lutz et al.  (1998),
and Lutz et al. (these proceedings) argue that the strength of the
7.7{\ts}$\mu$m PAH line/continuum ratio can be used to determine the fraction
of the infrared/submillimeter luminosity peak in SEDs that is due to a
starburst and an AGN respectively.  Using their arguments alone, two of the
four ULIGs in Table 2 (Mrk{\ts}231, Mrk{\ts}273) owe $\ga${\ts}50{\ts}\% of
their far-infrared/submillimeter luminosity to an AGN (e.g. Lutz et al., these
proceedings).  These authors also point out that there is evidence for
``coexistence of central AGN and circumnuclear star formation in a significant
fraction of (ULIGs)", and they note that ``the mid-infrared emitting regions
are highly obscured ($A_{\rm V} \sim 5-50$) for the screen case of $A_{\rm V}
\sim 50-1000$ for the fully mixed case)."  Large optical depths (i.e. $A_{\rm
V} > 1000$) along the line of sight toward the nuclei of ULIGs are indeed
suggested by millimeterwave interferometer measurements of molecular lines
(e.g. Scoville et al. 1991; Downes \& Solomon 1998; Sakamoto et al. 1999), as
well as from the weakness of far-infrared spectral lines such as the
158{\ts}$\mu$m C$^+$ line (J. Fischer, these proceedings) which indicate
optical depths of unity in ULIGs even at $\lambda \ga 100${\ts}$\mu$m !

High spatial resolution measurements of ULIGs in the mid-infrared can
potentially constrain the emitting size of the region responsible for the bulk
of the mid-infrared luminosity in ULIGs.  Soifer et al. (1999) report that the
bulk of the mid-infrared emission ($\Delta
\lambda${\ts}$\sim${\ts}10--25{\ts}$\mu$m) from Arp{\ts}220 comes from two very
compact regions centered on the radio nuclei, each with diameter
$\la${\ts}200{\ts}pc (FWHM).  Both a very compact starburst and an AGN can be
modeled to fit the data, however as pointed out by Soifer et al., it becomes
difficult to hide a $10^{12}${\ts}$L_\odot$ starburst in regions this small.

\subsubsection{Near-IR}

Strong emission lines such as Pa$\alpha$ and Pa$\beta$ as well as
high-excitation lines such as [Si VI] can potentially be used to distinguish
starbursts from AGN.  Even if the extinction is still two large along the line
of sight to observe emission lines from the central source, there is the
potential for observing scattered emission in polarized light.  Two of the four
ULIGs for which sensitive near-infrared spectra exist (IRAS{\ts}05189-2524,
Mrk{\ts}231) show clear evidence for polarized Seyfert{\ts}1 emission lines and
[Si VI] emission, further suggesting the presence of a strong AGN.

\subsubsection{Optical}

The use of standard line diagnostic diagrams (e.g. Veilleux \& Osterbrock 1987)
show that three of the five ULIGs have Seyfert spectra.  Both ULIGs with
Seyfert{\ts}1 optical and/or mid-infrared spectra have $L_{\rm BLR}/L_{bol}$
ratios nearly identical to the mean observed ratio for UV-excess QSOs,
consistent with the hypothesis that the bulk of the bolometric luminosity in
these infrared-dominated objects may indeed be due to a dust-enshrouded
UV-excess QSO.

\subsubsection{X-ray}

Like the radio, the X-ray luminosity is but a small fraction of the total
bolometric luminosity of all starbursts and most AGN, albeit usually a few
orders of magnitude in $\nu L \nu$ larger than for the radio, but still several
orders of magnitude less than the infrared luminosity of ULIGs.  It is not
clear that the observed X-ray spectrum can directly be related to the
bolometric luminosity of ULIGs.  However, by analogy with the mean X-ray
properties of QSOs or starbursts, it may at least be possible to say whether a
QSO-like X-ray nucleus is present.

The 2--10{\ts}keV luminosity of all five ULIGs in Table 2 is weak relative to
the strong far-infrared emission (e.g. Nakagawa et al. 1999); however, {\it the
X-ray luminosity of ULIGs is not weak in relation to RQQSOs when comparing the
X-ray luminosity to the luminosity in the optical} (e.g. Ogasaka et al. 1997;
Iwasawa 1999; Turner 1999).  Strong absorption appears to affect both the
optical/UV and soft X-rays in ULIGs, as might have been expected.

Absorption effects are minimized by considering only the hard X-ray emission,
$L_{\rm HX}${\ts}$\equiv${\ts}$L(5-10{\ts}{\rm keV})$, detected with {\it
ASCA}.  Three of the five ULIGs were detected in hard X-rays, all at a level
consistent with the mean hard X-ray luminosity observed for RQQSOs [Note:
Mrk{\ts}231 is compared here to the mean properties of BALQSOs which are
somewhat weaker in their observed $L_{\rm HX}$ than other RQQSOs (Turner
1999).]

\begin{figure}[htb]
\includegraphics{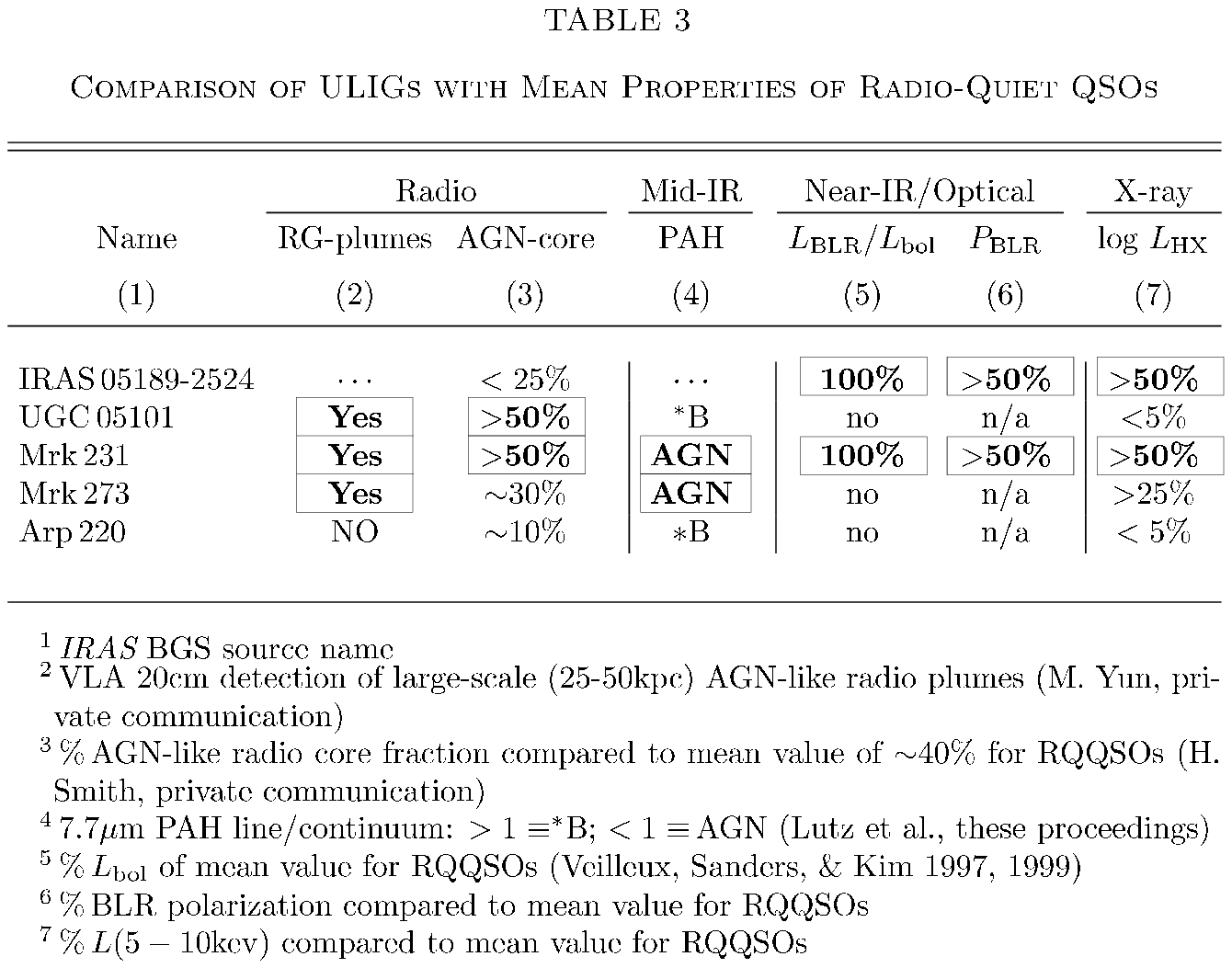}
\vspace{9cm}
\end{figure}

\subsubsection{Summary: Comparison with RQQSOs}
\vspace{0.01cm}

Table 3 compares the properties of the five nearest ULIGs with the {\it mean}
properties of RQQSOs, in order to try to answer the question --- Is there
evidence that $>${\ts}50{\ts}\% of the bolometric luminosity in
$>${\ts}50{\ts}\% of ULIGs is due to an AGN{\ts}?  Not surprisingly, perhaps,
is that the two ``warm'' ULIGs (IRAS{\ts}05189$-$2524, Mrk{\ts}231) show
substantial multiwavelength evidence for a dominant AGN, but equally important
is the fact that one additional object, Mrk{273}, is predicted by the starburst
camp to contain a dominant AGN (Lutz et al., these proceedings), and another,
UGC{\ts}05101, shows substantial evidence at radio wavelengths for harboring a
powerful AGN.

What is perhaps most surprising is that Arp{\ts}220 appears to be alone in this
small but well-studied group of the nearest ULIGs in it's absence of any clear
signature of a dominant AGN.  Rather than being the ``rosetta-stone'' for
ULIGs, Arp{\ts}220 may simply be the nearest such object, and perhaps one of
the most heavily obscured ULIGs as indicated by the extremely strong reddening
in the optical and near-infrared, as well as evidence for extremely strong
silicate absorption in the mid-infrared.

Have we shown that the answer to Topic 1 is ``an AGN'' ?  If you take the five
nearest ULIGs and adopt the bolometric luminosity indicator used by the
starburst camp (i.e. the 7.7{\ts}$\mu$m/continuum ratio) plus the bolometric
luminosity indicator used by the AGN camp (i.e.  $L_{\rm BLR}/L_{\rm bol}$),
then 4 of the 5 ULIGs in Table 3 are indeed dominated by an AGN.  However, our
decision to focus only on those objets that have been observed at the broadest
range of wavelengths, and with the highest resolution and sensitivity that
current X-ray satellites and radio interferometers can provide, has produced a
sample too small to statistically prove that $>${\ts}50{\ts}\% of {\it all}
ULIGs are dominated by AGN, thus our answer to Topic 1 must be ...

\setbox4=\vbox{\hsize 28pc \noindent \strut 
{\bf A1:\ Possibly.}

\hangindent=0.25in There is evidence from a detailed multiwavelength study of 
the 5 nearest ULIGs that $>${\ts}50\% of $L_{\rm bol}$
in $>${\ts}50\% of ULIGs is due to an heavily 
dust enshrouded AGN. \strut}

$$\boxit{\boxit{\box4}}$$

\section{Merger Sequence}

\setbox4=\vbox{\hsize 28pc \noindent\hangindent=0.25in \strut 
{\bf Q2:\ Do ULIGs follow a merger sequence from colliding disk galaxies 
with large bulges to ellipticals ?}\strut}  

$$\boxit{\box4}$$

\begin{figure}
\includegraphics{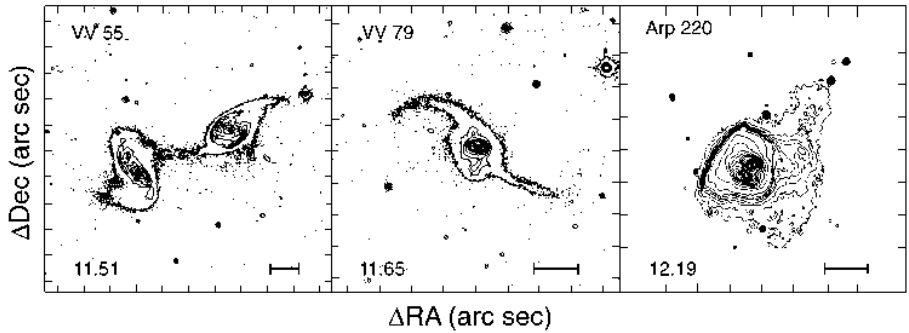}
\vspace{5cm}
\caption{A subsample of R-band images (Mazzarella et al. 1999) 
of luminous infrared galaxies from 
the {\it IRAS} RBGS (Sanders et al. 1999) 
that illustrate the strong interactions/mergers that are characteristic of 
nearly all objects with $L_{\rm ir} > 10^{11.3}{\ts}L_\odot$. 
The scale bar represents 10{\ts}kpc, tick marks are at 20$^{\prime\prime}$ 
intervals, and the infrared luminosity (log $L_{\rm ir}/L_\odot$) is 
indicated in the lower left corner of each panel.}
\end{figure}

Ground-based optical and near-infrared imaging of complete samples of the
brightest infrared galaxies clearly show that a substantial fraction of LIGs
are strongly interacting or merging spirals, and that the higher the luminosity
the more advanced is the merger (e.g. Joseph \& Wright 1985; Sanders, Surace,
\& Ishida 1999; Mazzarella et al. 1999).  Millimeterwave observations of have
shown these spirals to be rich in molecular gas  -- $M({\rm H}_2) \sim
10^9-3\times 10^{10}${\ts}$M_\odot$ (e.g. Sanders et al. 1988a; Mirabel et al.
1990; Sanders, Scoville, \& Soifer 1991) -- and that there is an increasing
central concentration of this gas with increasing infrared luminosity (Scoville
et al. 1991; Downes \& Solomon 1998). There is no clear evidence in favor of
early versus late-type spirals, only that they both typically appear to be
large (i.e. 0.5--2{\ts}$L^*$) and molecular gas-rich.  The three LIGs shown in
Figure 2 provide a coarse illustration of early, mid, and late type mergers
commonly represented in the complete samples of LIGs.  Comparison of these
images with numerical simulations (e.g. Barnes \& Hernquist 1992; Mihos \&
Hernquist 1994; C. Mihos, these proceedings) aids in allowing these objects to
be placed in a rough time sequence.

Nearly all ULIGs appear to be late-stage mergers (e.g. Sanders et al 1988a,b;
Melnick \& Mirabel 1990; Kim 1995; Murphy et al. 1996; Clements et al. 1996)
.  The large-scale ground-based images shown in the left panel of 
Figure 1 illustrate the largely overlapping disks that are seen in 
a {\it complete} sample of the nearest and best-studied ULIGs.  
Greater detail in the inner disks of these ULIGs is better revealed in 
the higher resolution ground-based images and {\it HST} images shown in the 
center and rightmost panels of Figure 1.  The mean lifetime for the ULIG 
phase, estimated from the observed mean separation and relative velocity 
of the merger nuclei, is $\sim${\ts}$1-2\times 10^8${\ts}yrs.

\begin{figure}
\includegraphics{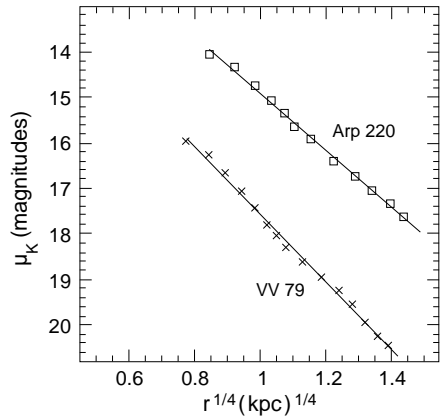}
\vspace{6cm}
\caption{K-band radial surface brightness profiles (Wright et al. 1990) for two
of the luminous infrared galaxies shown in Fig. 2 -- VV{\ts}79 (NGC{\ts}2623),
and Arp{\ts}220.  The straight line represents a normalized $r^{1/4}$-law
(deVaucouleurs) profile characteristic of elliptical galaxies.  The inner gap
corresponds to the lack of information at radii smaller than 1$^{\prime\prime}$
set by seeing.  More recent higher resolution K-band imaging of these galaxies
by NICMOS shows that a $r^{1/4}$-law continues to be a good representation of
the K-band radial profile over the range 0.1--5{\ts}kpc (Scoville et al.
1999).} \end{figure}

There is now substantial evidence that ULIGs are indeed elliptical galaxies
forming by merger-induced dissipative collapse (e.g. summary by Kormendy \&
Sanders 1992), including $r^{1/4}$-law brightness profiles [e.g. Schweizer
1982; Joseph \& Wright 1985; Wright et al. 1990 (see Figure 3); Kim 1995; Zheng
et al. 1999), newly-formed globular clusters (e.g. Surace et al. 1998), central
gas densities that are as high as stellar mass densities in the cores of giant
ellipticals (e.g. $\ga 10^2 M_\odot pc^{-3}$ at $r \la 0.5-1${\ts}kpc: Scoville
et al. 1991; Downes \& Solomon 1998), and powerful ``superwinds" that will
likely leave behind a largely dust free core (Heckman, Armus, \& Miley 1987;
Armus, Heckman, \& Miley 1989).

\setbox4=\vbox{\hsize 28pc \noindent \strut 
{\bf A2:\ YES, } 

\hangindent=0.25in ULIGs appear to represent an advanced stage in the merger of
two large spirals, where the merger remnant already has assumed a r$^{1/4}$-law
profile,

{\bf but} 

\hangindent=0.25in there is no evidence that only early-type 
spirals are involved, only that both progenitors be molecular gas-rich.\strut}  

$$\boxit{\boxit{\box4}}$$

\section{Precursors of QSOs}

\setbox4=\vbox{\hsize 28pc \noindent\hangindent=0.25in \strut 
{\bf Q3:\ Are ULIGs precursors of QSOs ?}\strut}  

$$\boxit{\box4}$$

If one chooses not to believe that ULIGs already harbor a dust enshrouded QSO
(i.e. an UV-excess AGN with $M_{\rm B} < -22.3$), then is there evidence that
they will become QSOs ?  It is probably a fair summary of the ``Green Tower''
view that at least 20--30{\ts}\% of ULIGs (i.e. the ``warm'' objects) already
harbor a bonifide QSO, and that a substantial fraction, if not all, of the cool
ULIGs have the {\it potential} to eventually become QSOs.

\begin{figure}
\includegraphics{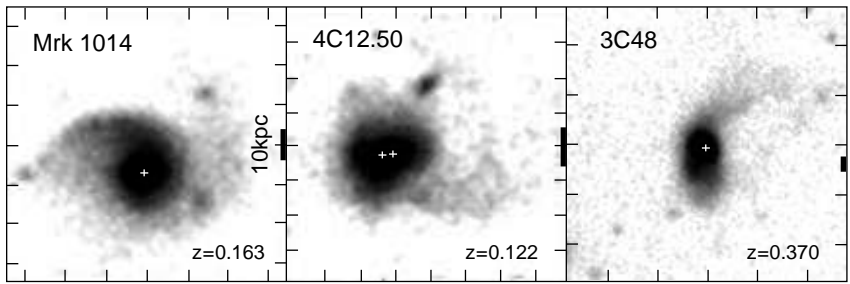}
\vspace{5cm}
\caption{Optical images of infrared-excess, optically selected QSOs, powerful
radio galaxies, and infrared selected QSOs (MacKenty \& Stockton 1984; Kim
1995; Stockton \& Ridgway 1991). The `$+$' sign indicates the position of
putative optical nuclei.  Tick marks are at 5$^{\prime\prime}$ intervals and
the scale bar represents 10{\ts}kpc.  All three objects exhibit strong nuclear
concentrations of molecular gas, with typically
$\sim${\ts}10$^{10}${\ts}$M_\odot$ concentrated at galactocentric radii
$\la${\ts}1{\ts}kpc (Sanders et al. 1988c; Mirabel, Sanders, \& Kaz\`es 1989;
Scoville et al. 1989).} \end{figure}

Of course there is no reason to believe that all ULIGs, once unshrouded, will
necessarily reach the optical/UV luminosity associated with QSOs -- they may
already have peaked in $L_{\rm bol}$ and/or some objects may simply be pure
starbursts that for some reason never choose to build/fuel a massive black
hole.  However, recent studies of the host galaxies of QSOs provide new
evidence for a plausible evolutionary connection between the ULIG phase and the
optical/UV excess QSO phase.  The mean and range of the H-band luminosity of
QSO hosts, $L_{\rm H} \sim 1-4{\ts}L^*$, reported by McLeod \& Rieke (1994) and
McLeod, Rieke, \& Storrie-Lombardi (1999) are remarkably similar to the H-band
luminosities of ULIGs (e.g. review by Sanders \& Mirabel 1996).  Also, it has
been known for some time that QSO hosts often exhibit tidal features indicative
of strong interactions/mergers [e.g. Stockton \& MacKenty 1983: MacKenty \&
Stockton 1984 (see the left panel in Figure 4)], and more recent {\it HST}
images of both radio-loud and radio-quiet QSOs show clear signs of large scale
tidal debris, circumnuclear knots, bars and rings (e.g. McLure et al. 1999)
similar to the inner structures seen in {\it HST} images of ULIGs (e.g. Surace
et al. 1998).  It is becoming easier to believe that QSO hosts are indeed
slightly more evolved stages of ULIG mergers.

\setbox4=\vbox{\hsize 28pc \noindent\hangindent=0.25in \strut 
{\bf A3:\ YES, }  

\hangindent=0.25in there is good evidence that $\sim${\ts}20--30\% of 
ULIGs are already dust-enshrouded QSOs, 

{\bf but} 

\hangindent=0.25in it is not at all clear that all ULIGs will obtain 
the optical/UV luminosity corresponding to bonafide QSOs 
(i.e. $M_{\rm B} < -22.3$). \strut} 

$$\boxit{\boxit{\box4}}$$
 
\section{Local Templates of High-z Mergers}

\setbox4=\vbox{\hsize 28pc \noindent\hangindent=0.25in \strut 
{\bf Q4:\ Are ULIGs local templates of the high luminosity tail of 
mergers at z = 1--4 ?}\strut}  

$$\boxit{\box4}$$

[Note:\ We have interpreted this question as asking whether there is evidence
that low-$z$ ULIGs are templates of high-$z$ ULIGs, leaving aside the larger
question of how high-$z$ ULIGs may be related to optically-selected high-$z$
objects (e.g. Lyman-break galaxies).]

There is now substantial evidence to suggest that the space density of ULIGs
evolves steeply with cosmic lookback time, and that ULIGs were much more common
at redshifts $z \sim${\ts}1--4.  In the mid- and far-infrared, the deepest
surveys carried out by {\it IRAS} (e.g. Hacking \& Houck 1987; Lonsdale \&
Hacking 1989; Gregorich et al. 1995; Kim \& Sanders 1998), and more recently
the deep surveys with {\it ISO} (e.g. Taniguchi et al. 1997; Kawara et al.
1998; Aussel et al. 1999; Puget et al. 1999) are consistent with number density
evolution as steep as $(1+z)^5$ out to $z \sim$1.  Within the past year,
submillimeter surveys with SCUBA on the JCMT (Smail, Ivison, \& Blain 1997;
Hughes et al. 1998; Barger et al. 1998; Eales et al. 1999) have revealed what
appears to be a substantial population of high-$z$ ULIGs (i.e. $z
\sim${\ts}1--4) , that are plausibly the high-$z$ extension of the low-$z$
ULIGs detected by {\it IRAS}.

It is still too early to tell whether all of the high-$z$ ULIGs detected by
{\it ISO} and SCUBA have properties similar to local ULIGs.  However, studies
of the few high-$z$ objects whose redshifts have been identified show that
these sources may indeed resemble their lower redshift counterparts more
closely than might at first have been assumed.  The two best studied sources
from the SCUBA sample of Smail et al.  (1998) are illustrative.
SMM{\ts}J02399$-$0136 at $z \sim 2.8$, with $L_{\rm ir} \ga 10^{13} L_\odot$,
is morphologically compact with an optical classification as a narrow-line
``type-2" AGN (Ivison et al. 1998; Ivison, these proceedings), and contains
$\sim${\ts}$10^{10.5} M_\odot$ of molecular gas (Frayer et al. 1998).
SMM{\ts}J14011+0252 at $z \sim 2.6$ with $L_{\rm ir} \sim 10^{12.3} L_\odot$
(Barger et al. 1999), is a strongly interacting/merger pair, with an
H{\ts}II-like optical spectrum and $\sim${\ts}$10^{10.7} M_\odot$ of molecular
gas (Frayer et al. 1999).  These two sources fit into the pattern exhibited by
ULIGs in the local Universe.  In particular their molecular gas masses, optical
luminosities, and optical morphologies are very similar to what is observed for
local ULIGs [see Sanders \& Mirabel (1996) for a review of local ULIGs].

\setbox4=\vbox{\hsize 28pc \noindent \strut 
{\bf A4:\ YES, }  

\hangindent=0.25in There is evidence that local ULIGs have properties 
(gas content, morphology, etc. ) similar to the identified infrared 
luminous sources at $z =${\ts}1--4, 

{\bf but} 

\hangindent=0.25in there is no reason to believe that all high-$z$, 
high-$L_{\rm ir}$ sources are as homogeneous as the low-$z$ 
ULIGs appear to be.\strut}  

$$\boxit{\boxit{\box4}}$$
 

\section{Concluding Remarks}

The Scientific Organizing Committee is to be commended for expanding this
debate to include a wider range of topics that are clearly important for the
study of ULIGs.  Both sides in the debate would likely agree that ULIGs
represent an extremely important stage in galaxy transformations (e.g. through
the building of ellipticals from merging spirals) and the accompanying metal
enrichment of the intergalactic medium via nuclear superwinds.  Both sides
would probably also agree that research on local ULIGs has become even more
important following the extremely exciting discoveries from {\it ISO} and SCUBA
that suggest that the far-infrared/submillimeter background radiation (as
measured by {\it COBE}) may indeed be resolved into a population of ULIGs at $z
\sim 1-4$, and that the far-infrared/submillimeter luminosity from these
high-$z$ ULIGs may dominate that attributed to star formation as seen in the
rest-frame optical/UV.

It also seems clear that most participants at this workshop believe that {\it
both} starbursts and AGN are important at varying levels to the luminosity
output of ULIGs, and that there is most likely an evolutionary connection
between the fueling of starbursts and the fueling of AGN.  Indeed, what better
time to fuel both a circumnuclear starburst and an AGN than when dumping
$\sim${\ts}$10^{10}{\ts}M_\odot$ of gas and dust into the central kiloparsec of
a merger remnant.  And if ULIGs indeed represent the building of massive bulges
during the merger of gas-rich spirals, then the recent discovery that all
massive ellipticals appear to contain massive black holes, where $M_{\rm MBH}$
is a constant fraction (i.e $\sim${\ts}0.006) of $M_{\rm bulge}$ (e.g. Kormendy
\& Richstone 1995; Magorrian et al. 1998) also suggests that the
building/fueling of a massive black hole is likely to be concurrent with the
ULIG phase.  And even if star formation is still favored as the dominant
luminosity source in most ULIGs, members of the starburst camp might consider
``affiliate membership" in the AGN camp simply by subscribing to the hypothesis
that most ULIGs may eventually evolve into optically-selected UV-excess QSOs.

The majority of workshop participants on both sides of this debate seemed
willing to accept the idea that ``warm'' ULIGs and those objects with the
highest infrared luminosities (e.g. log{\ts}$[L_{\rm ir}/L_\odot] \ga 12.6$)
are most likely to be powered by a dominant AGN.  Harder to accept was a
dominant role for AGN in the lower luminosity, cooler ULIGs like Arp{\ts}220,
but the fact that many Arp{\ts}220-like objects appear to have radio and hard
X-ray properties similar to the mean properties of RQQSOs (e.g. Mrk{\ts}273,
UGC{\ts}05101) suggests that dust obscuration could still mask a dominant AGN
in the majority of these objects as well.

\vspace{0.4cm} 

\noindent
{\bf Acknowledgments.}\ I am grateful to Jason Surace and Karen Teramura for
assistance in preparing the figures, to Aaron Evans, Jackie Fisher, Hagai
Netzer, Gene Smith, Sylvain Veilleux, and Min Yun for supplying data used in
this summary, to all of the members of the AGN-tower discussion group (the
``green team'') for their comments and suggestions, and to JPL contract no.
961566 for partial financial support.

\end{document}